\documentclass{iau} 
\usepackage{graphicx}
\usepackage{amssymb, amsmath}
\usepackage{subcaption}
\usepackage{rotating}
\usepackage{pdflscape}
\usepackage{xcolor}
\usepackage{ulem}

\def\kms{\,km\,s$^{-1}$}

\def\hb{{\sc{H}}$\beta$\/}

\def\feii{Fe{\sc{ii}}}
\def\rfe{R$_{\rm{FeII}}$}

\def\mbh{$M\mathrm{_{BH}}$}

\def\LLEdd{$L\mathrm{_{bol}}/L\mathrm{_{Edd}}$}

\def\RL{$R\mathrm{_{H\beta}}-L_{5100}$}

\def\zsun{Z$_{\odot}$}
\def\msun{M$_{\odot}$}

\def\rblr{$R\mathrm{_{BLR}}$}

\title[\feii{} emission in NLS1s] 
{\feii{} emission in NLS1s -- originating from denser regions with higher abundances?}

\author[Swayamtrupta Panda, Paola Marziani \& Bo\.zena Czerny]   
{Swayamtrupta Panda$^{1,2}$, Paola Marziani$^3$ \and Bo\.zena Czerny$^1$}

\affiliation{$^1$Center For Theoretical Physics, Polish Academy of Sciences, Al. Lotnik\'ow 32/46, 02-668 Warsaw, Poland \\ email: {\tt panda@cft.edu.pl} \\[\affilskip]
$^2$Nicolaus Copernicus Astronomical Center, Polish Academy of Sciences, ul. Bartycka 18, 00-716 Warsaw, Poland \\
$^3$ INAF-Astronomical Observatory of Padova, Vicolo dell'Osservatorio, 5, 35122 Padova PD, Italy}

\pubyear{2019}
\volume{356}  
\jname{IAU356: Nuclear Activity in Galaxies across Cosmic Time}
\editors{M. Povic et al., eds.}
\begin{document}

\maketitle

\begin{abstract}
The interpretation of the main sequence of quasars has become a frontier subject in the last years. The consider the effect of a highly flattened, axially symmetric geometry for the broad line region (BLR) on the parameters related to the distribution of quasars along their Main Sequence. We utilize the photoionization code CLOUDY to model the BLR, assuming `un-constant' virial factor with a strong dependence on the viewing angle. We show the preliminary results of the analysis to highlight the co-dependence of the Eigenvector 1 parameter, \rfe{} on the broad \hb{} FWHM (i.e. the line dispersion) and the inclination angle ($\theta$), assuming fixed values for the Eddington ratio (\LLEdd{}), black hole mass (\mbh{}) and spectral energy distribution (SED) shape. We consider four cases with changing cloud density (n$\rm{_{H}}$) and composition. Understanding the \feii{} emitting region is crucial as this knowledge can be extended to the use of quasars as distance indicators for Cosmology.\footnote{The project was partially supported by NCN grant no. 2017/26/\-A/ST9/\-00756 (MAESTRO  9) and MNiSW grant DIR/WK/2018/12. PM acknowledges the INAF PRIN-\-SKA 2017 program 1.05.01.88.\-04.}

\end{abstract}

\firstsection 
\section{Introduction}
The quasar main sequence contextualizes and eases the interpretation of classes of active galactic nuclei (AGN) whose origin has been debated for decades. An important class is the one of
 Narrow-Line Seyfert 1 (NLS1) galaxies which constitute a class of Type-1 active galaxies with ``narrow'' broad profiles. Their supermassive black holes (BH) are believed to have masses lower than the typical broad-line Seyfert galaxies. Black hole masses are estimated assuming that the line broadening is due to Doppler effect associated with the emitting gas motion with respect to the observer. In addition, the motions are believed to be predominantly virial (\cite[Peterson \& Wandel 1999]{peterson99}).  If the virial assumption is verified, the \mbh{} can be written as a function of (i) the radius of the broad line emitting region (BLR); and (ii) the FWHM of the emission lines emitted by gas whose motions are assumed virialized. The  BLR radius (\rblr{}) is derived via reverberation mapping (\cite[Peterson 1993]{peterson93}) i.e., by measuring the light-travel time from the central ionizing source to the line emitting medium. The line FWHM can be reliably measured from high S/N spectroscopy.

The BLR is a complex region, even if its physics is overwhelmingly driven by the process of photoionization. It cannot be characterized by a single quantity or number. The origin of different ionic species from this region and the advent of the reverberation mapping to probe more emission lines, have shown that the BLR is indeed stratified in terms of its density and  structure. Newer observations, such as of the Super-Eddington sources (\cite[see]{du2018} and references therein), have opened up a new field in the study of quasars. And one such immediate application is the use of these Super-Eddington sources, which are primarily NLS1s, as ``standardizable'' Eddington candles furthering the use of quasars in cosmology (see \cite[Marziani et al. 2019]{mar19}, \cite[Mart\'inez Aldama et al. 2019]{mlma19} and references therein). 

We address this aspect of the geometry of the quasars using photoionisation modelling with CLOUDY in the context of understanding better the main sequence of quasars (see \cite[Panda et al. 2019a]{Panda19a}, \cite[Panda et al. 2019b]{Panda19b} for more details). We focus on modelling the \feii{} emission in quasars as a function of the 7 key parameters -- (i) black hole mass (\mbh{}); (ii) Eddington ratio (\LLEdd{}); (iii) shape of the broad-band ionizing continuum (SED); (iv) mean cloud density (n$_{\rm{H}}$); (v) cloud metallicity; (vi) micro-turbulence; and (vii) H$\beta$ FWHM distribution. This multi-parameter space is then visualized as a function of the inclination angle ($\theta$) of the source wrt the observer. Here, we illustrate the results for \rfe{} estimated from a BLR cloud primarily as a function of the FWHM and $\theta$.  

\section{Method}

We assume a single cloud model where the density (n$_\mathrm{H}$) of the ionized gas cloud is varied from $10^{9}\; \mathrm{cm^{-3}}$ to $10^{13}\; \mathrm{cm^{-3}}$ with a step-size of 0.25 (in log-scale). We utilize the \textit{GASS10} model \cite[Grevesse et al. (2010)]{gass10} to recover the solar-like abundances and vary the metallicity within the gas cloud, going from a sub-solar type (0.1 Z$_{\odot}$) to super-solar (100 Z$_{\odot}$) with a step-size of 0.25 (in log-scale). The total luminosity of the ionizing continuum is derived assuming a value of the Eddington ratio (\LLEdd{}) and the respective value for the black hole mass (here, we assume an \LLEdd{} = 0.25 and a \mbh{} = 10$^8$ \msun{}). These values are appropriate for the part of Population A in spectral types. The shape of the ionizing continuum used here is taken from \cite[Korista et al. (1997)]{kor97}. The size of the BLR is estimated from the virial relation, assuming a black hole mass, a distribution in the viewing angle [0-90 degrees] and FWHM (for more details see \cite[Panda et al. 2019a]{Panda2019a}, \cite[Panda et al. 2019b]{Panda2019b}). The cloud column density (N$_\mathrm{H}$) is assumed to be $10^{24}\; \mathrm{cm^{-2}}$. 

The virial relation can be expressed as
\begin{equation}
    R_{BLR} = \frac{GM_{BH}}{f*FWHM^2} = \frac{4GM_{BH}\left[\kappa^2 + \sin^2\theta\right]}{\mathrm{FWHM}^2}
    \end{equation}

Substituting the values for the \mbh{} and $\kappa$ (= 0.1 that is consistent with a flat, keplerian-like gas distribution) in the virial relation, we have

\begin{equation}
    R_{BLR} \approx 5.31\times 10^{24}\left[\frac{0.01 + \sin^2\theta}{\mathrm{FWHM}^2}\right] \quad\quad\quad(\rm{in\;cm})
    \label{eq:3}
\end{equation}

\section{Results and Conclusions}
\begin{figure}
\hfill
\vfill
\begin{turn}{90}
\begin{minipage}[c][\textwidth][c]{1.25\textheight}
\begin{subfigure}[h]{.475\linewidth}
\includegraphics[width=\linewidth]{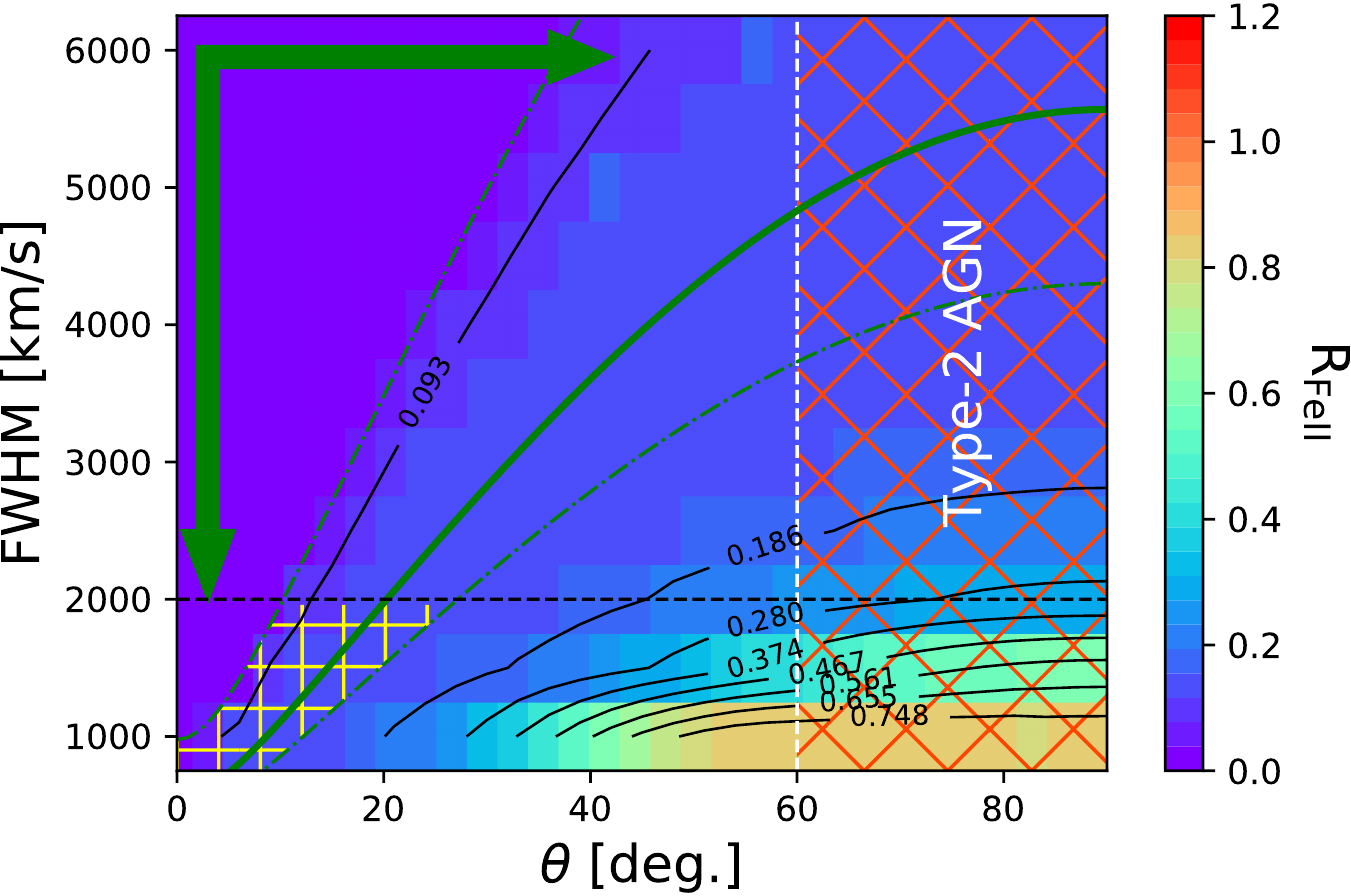}
\end{subfigure}\hfill
\begin{subfigure}[h]{.475\linewidth}
\includegraphics[width=\linewidth]{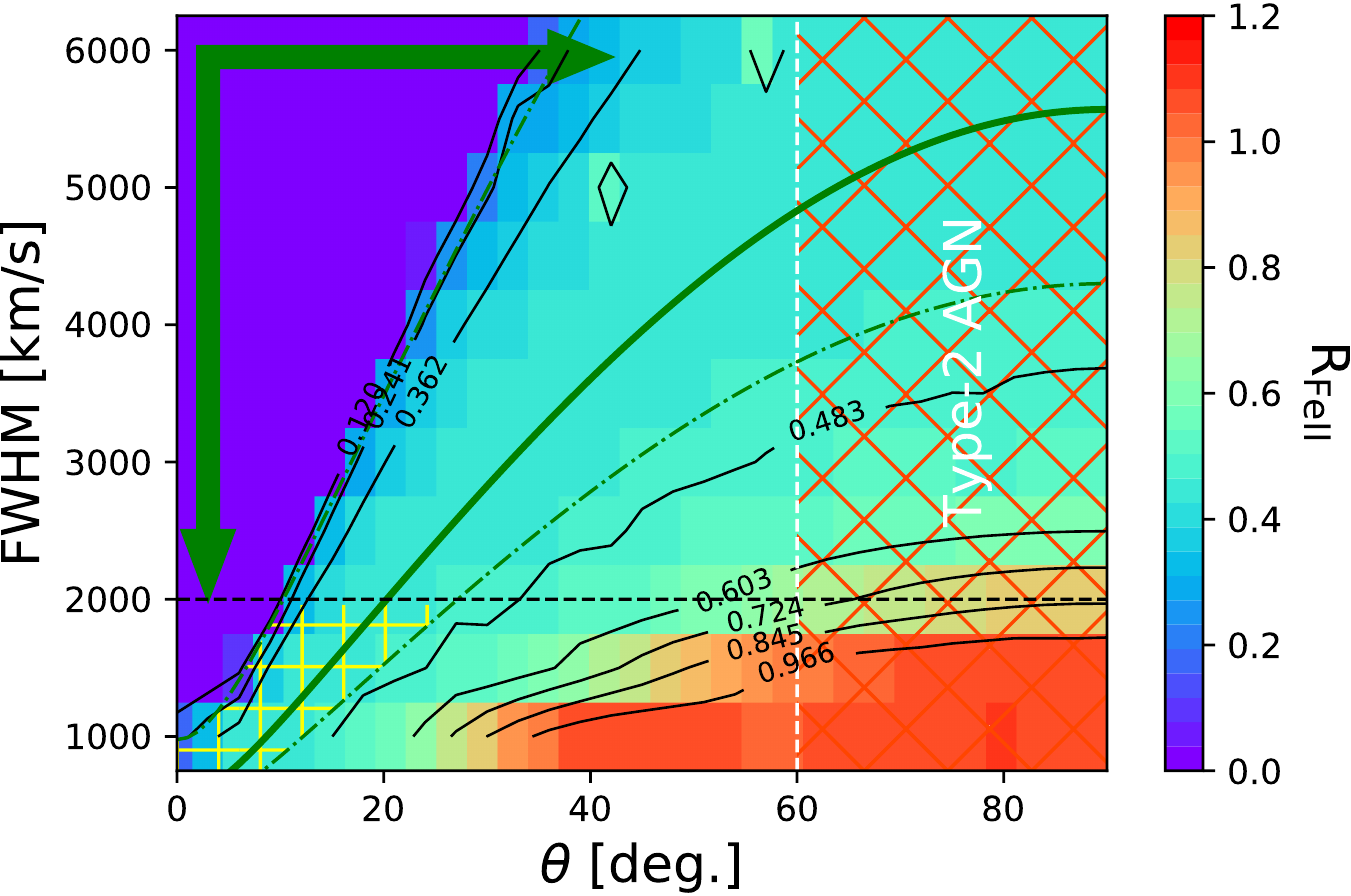}
\end{subfigure}
\caption{2D histogram showing the dependence of the parameter \rfe{} on the \hb{} FWHM and the inclination angle ($\theta$). The \hb{} FWHM ranges from 1000 \kms{} to 6000 \kms{} with a step size of 500 \kms{}. Similarly, the $\theta$ values range from 0$^{\rm{o}}$-90$^{\rm{o}}$ with a step size of 3$^{\rm{o}}$. The black hole mass is assumed to be 10$^8$\msun{}, the Eddington ratio, \LLEdd{} = 0.25 and a SED consistent with \cite[Korista et al. (1997)]{kor97} is used. The value of the $\kappa$ = 0.1 consistent with a flat, keplerian-like gas distribution around the central supermassive black hole. The mean cloud density (n$_{\rm{H}}$) is 10$^{10}$ cm$^{-3}$ and the column density is 10$^{24}$ cm$^{-2}$. The white dashed line marks the upper limit on the $\theta$ consistent with Type-1 sources. The hatched region marks the Type-2 AGN zone beyond $\theta$ = 60$^{\rm{o}}$. The solid green line traces the FWHM-$\theta$ for the \rblr{} estimated from the standard \RL{} relation, and the dot-dashed green lines correspond to the 1$\sigma$ scatter in the \RL{} relation. The yellow patch for FWHM $\lesssim 2000$ \kms{} and within the 1$\sigma$ scatter around the standard \RL{} relation marks the zone of acceptance for the NLS1s (typical Population A type sources). The solid green arrows point in the direction of increasing \rblr{} based on the virial relation. \textbf{LEFT}: at solar abundance (\zsun{}); \textbf{RIGHT}: at 10\zsun{}.}
\label{fig1}
\end{minipage}
\end{turn}
\end{figure}

In the left panel of Figure \ref{fig1}, we assume the mean cloud density (n$_\mathrm{H}$) at 10$^{10}$ $\mathrm{cm^{-3}}$. The peak of the \rfe{} ($\sim$0.8415) is located at $\sim$ 60$^{\rm{o}}$ for FWHM = 1000 \kms{}. Within the realms of Type-1 AGNs, i.e., $\theta \lesssim 60^{\rm{o}}$, \rfe{} $\propto \theta$. On the other hand, \rfe{} is inversely related to the FWHM. From the virial relation we have, \rblr{} $\propto \frac{1}{FWHM^2}$. This implies, \rfe{} $\propto \sqrt{\mathrm{R_{BLR}}}$. In other words, increasing FWHM decreases \rblr{} which means higher radiation flux on the cloud that leads to depletion in \feii{} emission. Hence, \rfe{} decreases. Increasing the metallicity from solar (\zsun{}) to 10\zsun{} shifts the peak of the \rfe{} to $\sim$ 81$^{\rm{o}}$ still for the case with FWHM = 1000 \kms{}. Within the limits of $\theta \lesssim 60^{\rm{o}}$, the maximum value of \rfe{} is at $\theta \sim 45^{\rm{o}}$ (\rfe{} $\sim$ 1.0835). Trends of \rfe{} wrt $\theta$ and FWHM respectively, remain consistent to the previous case (at solar abundance).

\begin{figure}
\hfill
\vfill
\begin{turn}{90}
\begin{minipage}[c][\textwidth][c]{1.25\textheight}
\begin{subfigure}[h]{.475\linewidth}
\includegraphics[width=\linewidth]{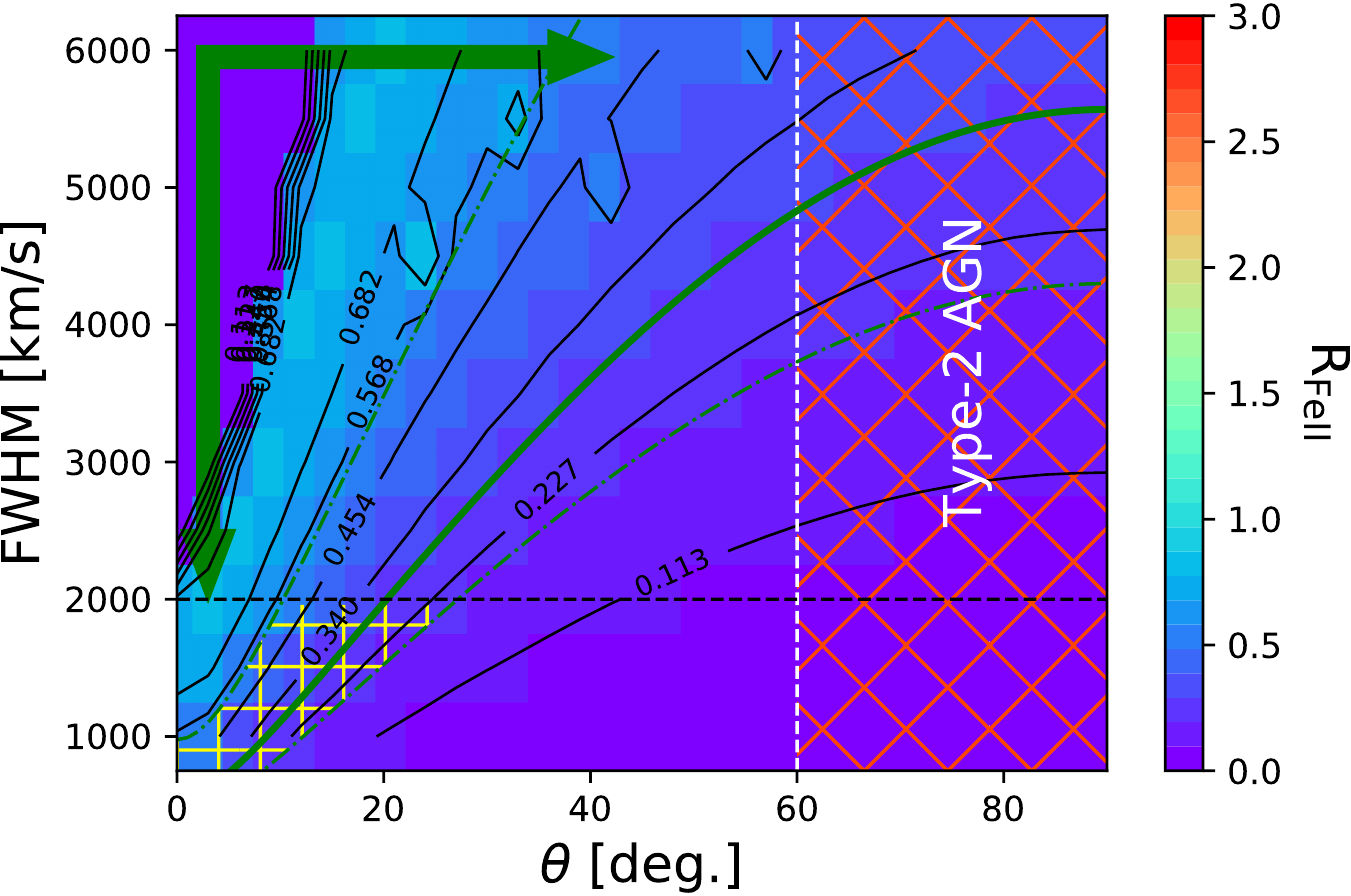}
\end{subfigure}\hfill
\begin{subfigure}[h]{.475\linewidth}
\includegraphics[width=\linewidth]{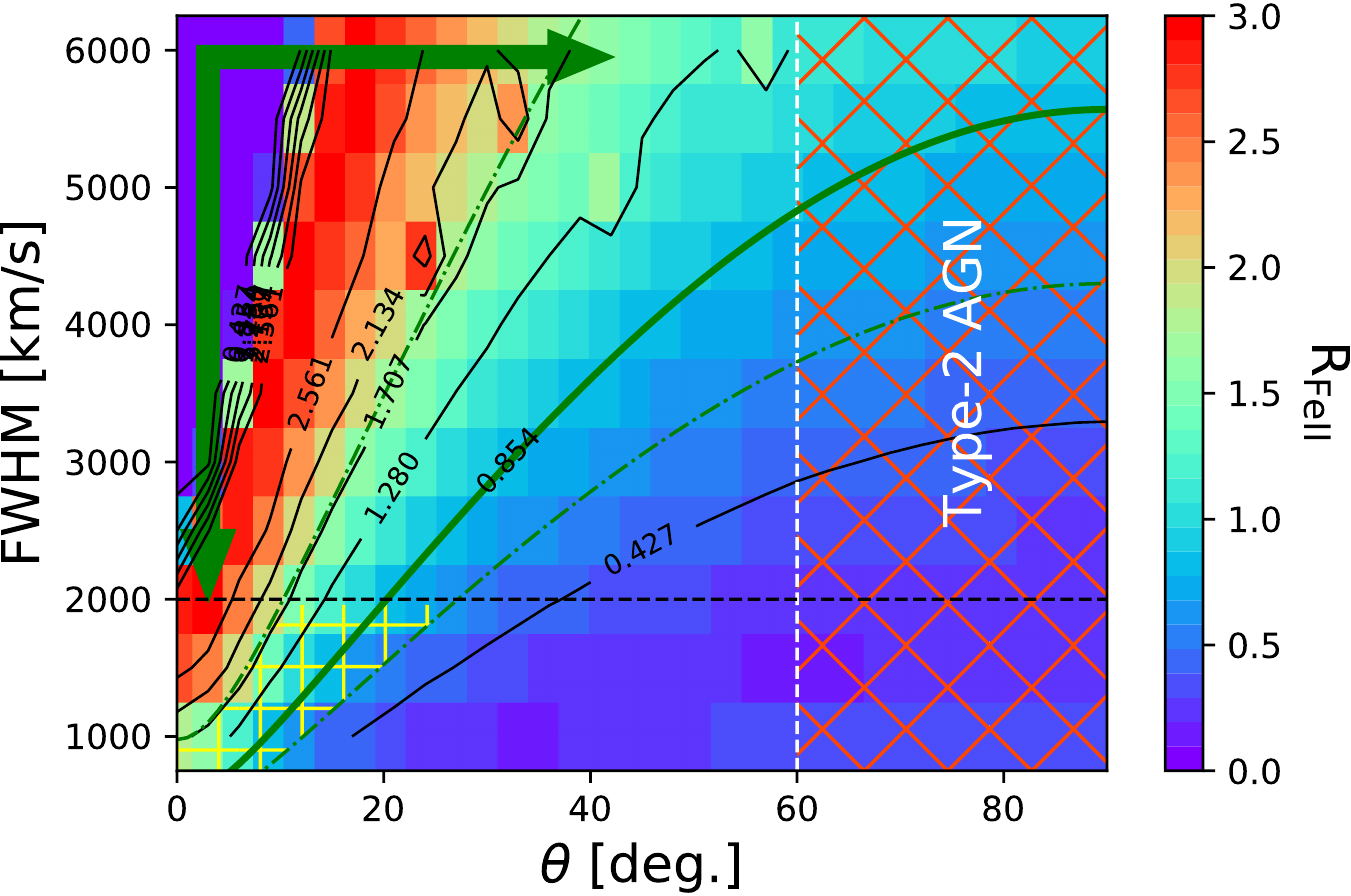}
\end{subfigure}
\caption{Same as Figure \ref{fig1}. The mean cloud density (n$_{\rm{H}}$) is increased to 10$^{12}$ cm$^{-3}$. \textbf{LEFT}: at solar abundance (\zsun{}); \textbf{RIGHT}: at 10\zsun{}.}
\label{fig2}
\end{minipage}
\end{turn}
\end{figure}

In Figure \ref{fig2}, we increase the value of n$_\mathrm{H}$ from 10$^{10}$ to 10$^{12}$ $\mathrm{cm^{-3}}$, keeping the other parameters exactly the same as before. The current value of density, i.e. 10$^{12}$ $\mathrm{cm^{-3}}$, is consistent with previous works related to the study of the main sequence of quasars (see \cite[Panda et al. 2018]{Panda18} and references therein). This change in the density changes the picture significantly. In the left panel of Figure \ref{fig2}, the peak value moves along one of the contour lines and within 2000 \kms{} $\lesssim$ FWHM $\lesssim$ 6000 \kms{}, the peak moves from $\sim 3^{\rm{o}}$ (for $\sim$2000 \kms{}) to $\sim 18^{\rm{o}}$ (for $\sim$6000 \kms{}). These peak values of \rfe{} remain at 0.77$\pm$0.01. The ionization parameter (\textit{U}) also changes accordingly. For instance, considering the \rblr{} from Eq. \ref{eq:3} at FWHM=2000\kms{} and at $\theta$= $3^{\rm{o}}$, gives \textit{U}$\sim$0.23 which in the low density case (10$^{10}$ $\mathrm{cm^{-3}}$) corresponds to a very high value, i.e. \textit{U}$\sim$23! 

Hence, the ionisation parameter governs the appearance of the plots.  A considerable region in the Figures \ref{fig1} and \ref{fig2} falls in  ``zones of avoidance'' where $U$\ is either too high or too low to sustain significant FeII emission (dark blue areas in the Figures). 
   In the case with the high density the peak emission is very close to the allowed zone by the \RL{} relation (the region within the 1$\sigma$ scatter of the \RL{} is shown with green dashed lines in Figures \ref{fig1} and \ref{fig2}). This implies that the NLS1s that are  high \feii{} emitters need to have a high density to boost their \feii{}. \rfe{} is even more enhanced in case of higher  metallicity. However, along the \RL{} line \rfe{} is constant, as expected since in this case all parameters affecting \feii{} intensity are set to a fixed value in our model. These results require further analysis which will be presented in a subsequent paper.

\end{document}